\documentclass[11pt,twoside]{article} %TS01-3 %TS01-3 %TS22-2
\usepackage{cspm-asp2006}
\usepackage{epsfig,graphicx,natbib,url}  %RR additions
\usepackage{lscape} %%\usepackage{hyperref}
\begin{document}
\setcounter{page}{49}

\markboth{Carlsson}{Modeling the Solar Chromosphere}
\title{Modeling the Solar Chromosphere}
\author{Mats Carlsson}   %%% Fill in author names
\affil{Institute of Theoretical Astrophysics, University of Oslo, Norway}    %%% Fill in author affiliations
\altaffiltext{1}{also at Center of Mathematics for Applications, University of Oslo,  Norway}    %%% Fill in author affiliations

\begin{abstract} %%% Abstract to run on from here.
Spectral diagnostic features formed in the solar chromosphere are few
and difficult to interpret --- they are neither formed in the
optically thin regime nor in local thermodynamic equilibrium (LTE).
To probe the state of the chromosphere, both from observations and
theory, it is therefore necessary with modeling.
I discuss both traditional semi-empirical modeling, numerical
experiments illustrating important ingredients necessary for a
self-consistent theoretical modeling of the solar chromosphere and the
first results of such models.
% forward modeling necessary
% semi-empirical modeling
% advances in modeling
% emphasis on basic philosophy behind theoretical modeling
\end{abstract}

%%%%%%%%%%%%%%%%%%%%%%%%%%%%%%%%%%%%%%%%%%%%%%%%%%%%%%%%%%%%%%%%%%%%%%%%%%%%
\section{Introduction}

% revised version of the SacPeak writeup
%
My keynote talk was similar in content to a recent talk at a
Sacramento Peak workshop celebrating the 70th birthday of Robert
F. Stein. This written version builds to a large extent on that
writeup \citep{mc-Carlsson2006}, but it is updated and some sections
have been expanded.

% what is the chromosphere
%  not self evident: 
%  different alternatives:
%   temperature rise - region above temperature minimum and up to TR
%   non-radiative/non-convective input of energy needed
%   historical: colour band - refer to Judge
Before discussing models of the solar chromosphere it is worthwhile
discussing the very definition of the term ``chromosphere''. The name
comes from the Greek words ``$\chi\rho\omega\mu\alpha$'' (color) and
``$\sigma\varphi\alpha\iota\rho\alpha$'' (ball) alluding to the
colored thin rim seen above the lunar limb at a solar eclipse. The
color comes mainly from emission in the Balmer H$\alpha$ 
line.  This is thus one possible definition --- the chromosphere is
where this radiation originates. At an eclipse this region has a sharp
lower edge, the visible limb, but a fuzzy upper end with prominences
protruding into the corona. The nature of this region is difficult to
deduce from eclipse observations since we see this region edge on
during a very short time span and we have no way of telling whether it
is homogeneous along the line of sight or very inhomogeneous in space
and time. It was early clear that the emission in H$\alpha$ must mean
an atmosphere out of radiative equilibrium --- without extra heating
the temperature will not be high enough to have enough hydrogen atoms
excited to the lower or upper levels of the transition. Early models
were constructed to explain observations in H$\alpha$ and in
resonance lines from other abundant elements with opacity high enough
to place the formation in these regions even in center-of-disk
observations (lines like the H and K resonance lines from singly
ionized calcium). These early models were constructed assuming one
dimensional plane-parallel geometry and they resulted in a temperature
falling to a minimum around 4000\,K about 500\,km above the visible
surface, a temperature rise to 8000\,K at a height of about 2000\,km and
then a very rapid temperature rise to a million degree corona. These
plane-parallel models have led to a common notion that there is a more
or less homogeneous, plane-parallel region between these heights that
is hotter than the temperature minimum. In such a picture the
chromosphere may be defined as a region occupying a given height range
(e.g.\ between 500 and 2000\,km height over the visible surface) or a
given temperature range. We may also use physical processes for our
definition: the chromosphere is the region above the photosphere where
radiative equilibrium breaks down and hydrogen is predominantly
neutral (the latter condition giving the transition to the
corona). This discussion shows that there is no unique definition of
the term ``chromosphere'', not even in a one-dimensional, static
world. It is even more difficult to agree on a definition of the
``chromosphere'' that also encompasses an inhomogeneous, dynamic
atmosphere.

% Different types of models: 
%  semi-empirical
%  theoretical

As mentioned above, the first models of the chromosphere were
constructed with a large number of free parameters to match a set of
observational constraints. Since some equations are used to restrict
the number of free parameters (not all hydrodynamical variables at all
points in space and time are determined empirically) we call this
class of models {\em semi-empirical} models.  Typically one assumes
hydrostatic equilibrium and charge conservation but no energy equation.
The temperature as function of height is treated as a free function to be
determined from observations. 
In the other main class of models one tries to
minimize the number of free parameters by including an energy equation. Such
theoretical models have been very successful in explaining radiation
from stellar photospheres with only the effective temperature, acceleration
of gravity and abundances as free parameters. In the chromosphere, an
additional term is needed in the energy equation --- e.g.\ energy deposition by
acoustic shocks or energy input in connection with magnetic fields (e.g.\
currents or reconnection).

% chromospheric ``heating''

It is thus clear from observations that the chromosphere is not in
radiative equilibrium --- there is a net radiative loss. This loss has
to be balanced by an energy deposition, at least averaged over a long
enough time span, if the atmosphere is to be in equilibrium. This is
often called the problem of chromospheric ``heating''.  It is important
to bear in mind, though, that the radiative losses may be balanced by
a non-radiative energy input without an increase in the average
temperature. The term ``chromospheric heating'' may thus be misleading
since it may be interpreted as implying that the average temperature
is higher than what is the case in a radiative equilibrium
atmosphere. In the following we will use the term ``heating'' in a more
general sense: a source term in the energy equation, not
necessarily leading to an increased temperature.

% restrict ourselves to Solar (no spots?)

Chromospheric heating is needed not only for the quiet or average Sun but
also in active regions, sunspots and in the outer atmospheres of many 
other stars. I will in the following mainly discuss the quiet Sun case.

% outline

The outline of this paper is as follows: In Section 2 we discuss
semi-empirical models of the chromosphere. In Section 3 we discuss
theoretical models; first we elaborate on 1D hydrodynamical models,
then we discuss the role of high frequency acoustic waves for the
heating of the chromosphere and finally we describe recent attempts to
model the chromosphere in 3D including 
the effects of magnetic fields. 

\section{Semi-empirical Models}

% basic philosophy
%  inversion codes vs forward modeling
%  unknown physics
%  assumptions

Semi-empirical models can be characterized by the set of observations
used to constrain the model, the set of physical approximations
employed and the set of free parameters to be determined.
Spectral diagnostics used to constrain chromospheric models must have
high enough opacity to place the formation above the photosphere. The
continuum in the optical part of the spectrum is formed in the
photosphere so the only hope for chromospheric diagnostics lies in
strong spectral lines in this region of the spectrum. Candidates are
resonance lines of dominant ionization states of abundant elements and
lines from excited levels of the most abundant elements (hydrogen and
helium). Most resonance lines are in the UV but the resonance lines of
singly ionized calcium (Ca\,II), called the H and K lines, fulfill our
criteria. These lines originate from the ground state of Ca\,II, the
dominant ionization stage under solar chromospheric conditions, and
the opacity is therefore given by the density directly and the optical
depth is directly proportional to the column mass (i.e.\ to the total
pressure in hydrostatic equilibrium).
Also the source function has some coupling to local conditions even
at quite low densities (in contrast to the strongly scattering resonance
lines of neutral sodium).
Other chromospheric diagnostic lines in the optical region are
the hydrogen Balmer lines and the helium 1083\,nm line. They all originate
from highly excited levels and thus have very temperature sensitive 
opacity. The population of He\,1083 is also set by recombination such that
its diagnostic potential is very difficult to exploit.
With the advent of space based observatories, the
full UV spectral range was opened up. Continua shortward of the opacity
edge from the ground state of neutral silicon at 152\,nm are formed above
the photosphere and can be used to constrain chromospheric models.
%  VAL
Together with observations in Ly-$\alpha$, such UV
continuum observations were used by \citet*{mc-VALI,mc-VALII,mc-VALIII}
in their seminal series of papers on the solar chromosphere. 
The VAL3 paper \citep{mc-VALIII} is one of the most cited papers in solar
physics (1072 citations in ADS at the time of writing) and the abstract
gives a very concise description of the models and the principles behind
their construction:
``The described investigation is concerned with the solution of the
non-LTE optically thick transfer equations for hydrogen, carbon, and
other constituents to determine semi-empirical models for six
components of the quiet solar chromosphere. For a given
temperature-height distribution, the solution is obtained of the
equations of statistical equilibrium, radiative transfer for lines and
continua, and hydrostatic equilibrium to find the ionization and
excitation conditions for each atomic constituent. The emergent
spectrum is calculated, and a trial and error approach is used to
adjust the temperature distribution so that the emergent spectrum is
in best agreement with the observed one. The relationship between
semi-empirical models determined in this way and theoretical models
based on radiative equilibrium is discussed by Avrett (1977). Harvard
Skylab EUV observations are used to determine models for a number of
quiet-sun regions.''

The VAL3 models are thus characterized by them using Ly-$\alpha$ and
UV-continuum observations for observational constraint, hydrostatic
equilibrium and non-LTE statistical equilibrium in 1D as physical
description and temperature as function of height as free function.
To get a match with observed line-strengths, a depth-dependent 
microturbulence was also determined and a corresponding turbulent
pressure was added. The number of free parameters to be determined
by observations is thus large --- in principle the number of depth-points 
per depth-dependent free function (temperature and microturbulence).
In practice the fitting was made by trial and error and only rather
smooth functions of depth were tried thus decreasing the degrees
of freedom in the optimization procedure.

The models have a minimum temperature around 500\,km above the visible
surface (optical depth unity at 500\,nm), a rapid temperature rise outwards
to about 6000\,K at 1000\,km height and thereafter a gradual temperature increase
to 7000\,K at 2000\,km height with a very rapid increase from there to coronal
temperatures.

%   problem with Tmin, MACKKL

The Ca\,II lines were not used in constraining the VAL3 models and 
the agreement between the model representing the average quiet Sun,
VAL3C, and observations of these lines was not good. An updated model with a different
structure in the temperature minimum region was published
in \citet{mc-MACKKL} (where the main emphasis was on similarly constructed
semi-empirical models for sunspot atmospheres).

%  FAL ambipolar diffusion, theoretical TR

A peculiar feature with the VAL models was a temperature plateau introduced
between 20000 and 30000\,K in order to reproduce the total flux in the
Lyman lines. This plateau was no longer necessary in the FAL models where the semi-empirical
description of the transition region temperature rise was replaced by
the balance between energy flowing down  from the corona (conduction and
ambipolar diffusion) and radiative losses 
\citep{mc-Fontenla+Avrett+Loeser1990,mc-Fontenla+Avrett+Loeser1991,mc-Fontenla+Avrett+Loeser1993}. 

One goal of semi-empirical models is to obtain clues as to the non-radiative
heating process. From the models it is possible to calculate the amount of
non-radiative heating that is needed to sustain the model structure. For
the VAL3C model this number is 4.2~kW\,m$^{-2}$ with the dominant radiative
losses in lines from Ca\,II and Mg\,II, with Ly-$\alpha$ taking over
in the topmost part.

%  Anderson, Athay

The models described so far do not take into account the effect of the very many
iron lines. This was done in modeling by \citet{mc-Anderson+Athay1989}. Instead of
using the temperature as a free parameter and observations as the 
constraints, they adjusted the non-radiative heating function
until they obtained the same temperature structure as in the VAL3C model
(arguing that they would then have an equally good fit to the observational constraints
as the VAL3C model). The
difference in the physical approximations is that they included line blanketing
in non-LTE from millions of spectral lines. The radiation losses are dominated
by Fe\,II, with Ca\,II, Mg\,II, and H playing important, but secondary, roles.
The total non-radiative input needed to balance the radiative losses is three
times higher than in the VAL3C model, 14 kW\,m$^{-2}$.

%  more semi-empirical models, CO problem

The VAL3 and FAL models show a good fit to the average (spatial and
temporal) UV spectrum but fail to reproduce the strong lines from
CO. These lines show very low intensities in the line center when
observed close to the solar limb, the radiation temperature is as low
as 3700\,K \citep{mc-Noyes+Hall1972, mc-Ayres+Testerman1981,
  mc-Ayres+Testerman+Brault1986, mc-Ayres+Wiedemann1989,
  mc-Ayres+Brault1990}. If the formation is in LTE this translates
directly to a temperature of 3700\,K in layers where the inner wings of
the H and K lines indicate a temperature of 4400\,K. The obvious
solution to the problem is that the CO lines are formed in non-LTE
with scattering giving a source function below the
Planck-function. Several studies have shown that this is not the
solution --- the CO lines are formed in LTE
\citep[e.g.][]{mc-Ayres+Wiedemann1989, mc-Uitenbroek2000}. The model M\_CO
constructed to fit the CO-lines \citep{mc-Avrett1995} give too low UV
intensities. One way out is to increase the number of free parameters
by abandoning the 1D, one-component, framework and construct a two
component semi-empirical atmosphere. The COOLC and FLUXT atmospheric
models of \citet{mc-Ayres+Testerman+Brault1986} was such an attempt where a filling
factor of 7.5\% of the hot flux tube atmosphere FLUXT and 92.5\% of
the COOLC atmosphere reproduced both the H and K lines and the
CO-lines. The UV continua, however, are overestimated by a factor of
20 \citep{mc-Avrett1995}. A combination of 60\% of a slightly cooler
model than M\_CO and 40\% of a hot F model  provides
a better fit \citep{mc-Avrett1995}. Another way of providing enough
free parameters for a better fit is to introduce an extra force in
the hydrostatic equilibrium equation providing additional support
making possible a more extended atmosphere. With this extra free
parameter it is possible to construct a 1D temperature structure
with a low temperature in the right place to reproduce the
%!! have added "near-limb observations of the" to clarify
near-limb observations of the CO lines and a
sharp temperature increase to give enough intensity in the UV
continua \citep{mc-Fontenla2007}.

%  free parameters - good fit not necessarily proof of realism

A word of caution is needed here. Semi-empirical models are often
impressive in how well they can reproduce observations. This is,
however, not a proper test of the realism of the models since the
observations have been used to constrain the free parameters. The
large number of free parameters (e.g., temperature as function of
height, microturbulence as function of height and angle, non-gravitational
forces) may hide
fundamental shortcomings of the underlying assumptions (e.g.,
ionization equilibrium, lateral homogeneity, static solution). It is
not obvious that the energy input required to sustain a model that
reproduces time-averaged intensities is the same as the mean energy
input needed in a model that reproduces the time-dependent intensities
in a dynamic atmosphere.  Semi-empirical modeling may give clues as
to what processes may be important but we also need to study these
underlying physical processes with fewer free parameters. This is the
focus of theoretical models.

\section{Theoretical Models}

%   studies of physical principles, cartoon models, numerical experiments
%   comprehensive physics
%    boundary from observations - compare directly
%    complete system, self-driven: statistical comparison

In contrast to semi-empirical models theoretical models include an
energy equation. To model the full 3D system with all physical
ingredients we know are important for chromospheric conditions is
still computationally prohibitive --- various approximations have to
be made. In one class of modeling one tries to illustrate basic
physical processes without the ambition of being realistic enough to
allow detailed comparison with observations. Instead the aim is to
fashion a basic physical foundation upon which to build our
understanding.  The other approach is to start with as much realism as
can be afforded.  Once the models compare favourably with
observations, the system is simplified in order to enable an
understanding of the most important processes. I here
comment on both types of approaches.

\subsection{1D radiation hydrodynamic simulations}

Acoustic waves were suggested to be the agent of non-radiative energy
input already by \citet{mc-Biermann1948} and
\citet{mc-Schwarzschild1948}.  Such waves are inevitably excited by
the turbulent motions in the convection zone and propagate outwards,
transporting mechanical energy through the photospheric layers into
the chromosphere and corona. Due to the exponential decrease of
density with height, the amplitude of the waves increases and they
steepen into shocks. The theory that the dissipation of shocks heats
the outer atmosphere was further investigated by various authors, see
reviews by \citet{mc-Schrijver1995, mc-Narain+Ulmschneider1996}.

% Ulmschneider et al

% Carlsson, Stein 1992-2002 

% description of simulations

In a series of papers,
\citet{mc-Carlsson+Stein1992,mc-Carlsson+Stein1994,mc-Carlsson+Stein1995,mc-Carlsson+Stein1997,
mc-Carlsson+Stein2002} have explored the effect of acoustic waves on
chromospheric structure and dynamics. The emphasis of this modeling
was on a very detailed description of the radiative processes and on
the direct comparison with observations.  The full non-LTE rate
equations for the most important species in the energy balance
(hydrogen, helium and calcium) were included thus including the
effects of non-equilibrium ionization, excitation, and radiative
energy exchange on fluid motions and the effect of motion on the
emitted radiation from these species. To make the calculations
computationally tractable, the simulations were performed in 1D and
magnetic fields were neglected. To enable a direct comparison with
observations, acoustic waves were sent in through the bottom boundary
with amplitudes and phases that matched observations of Doppler shifts
in a photospheric iron line.

% summary of results

These numerical simulations of the response of the chromosphere to
acoustic waves show that the Ca\,II profiles can be explained by
acoustic waves close to the acoustic cut-off period of the
atmosphere. The simulations of the behaviour of the Ca\,II\,H line
reproduce the observed features to remarkable detail. The simulations
show that the three minute waves are already present at photospheric
heights and the dominant photospheric disturbances of five minute
period only play a minor modulating role
\citep{mc-Carlsson+Stein1997}.  The waves grow to large amplitude
already at 0.5 Mm height and have a profound effect on the
atmosphere. The simulations show that in such a dynamic situation it
is misleading to construct a mean static model
\citep{mc-Carlsson+Stein1994,mc-Carlsson+Stein1995}. It was even
questioned whether the Sun has an average temperature rise at
chromospheric heights in non-magnetic regions
\citep{mc-Carlsson+Stein1995}. The simulations also confirmed the
result of \citet{mc-Kneer1980} that ionization/recombination
timescales in hydrogen are longer than typical hydrodynamical
timescales under solar chromospheric conditions.  The hydrogen
ionization balance is therefore out of equilibrium and depends on the
previous history of the atmosphere. Since the hydrogen ionization
energy is an important part of the internal energy equation, this
non-equilibrium ionization balance also has a very important effect on
the energetics and temperature profile of the shocks
\citep{mc-Carlsson+Stein1992, mc-Carlsson+Stein2002}.
\citet{mc-Kneer1980} formulated this result as strongly as ``Unless
confirmed by consistent dynamical calculations, chromospheric models
based on the assumption of statistical steady state should be taken as
rough estimates of chromospheric structure.''

% what does not fit
% Judge et al

Are observations in other chromospheric diagnostics than the Ca\,II
lines consistent with the above mentioned radiation-hydrodynamic
simulations of the propagation of acoustic waves? The answer is
``No''. The continuum observations around 130\,nm are well matched by
the simulations \citep{mc-Judge+Carlsson+Stein2003} but continua
formed higher in the chromosphere have higher intensity in the
observations than in the simulations \citep{mc-Carlsson+Stein2002c}.
The chromospheric lines from neutral elements in the UV range are
formed in the mid to upper chromosphere. They are in emission at all
times and at all positions in the observations, and they show stronger
emission than in the simulations.

The failure of the simulations to reproduce diagnostics formed in the
middle to upper chromosphere gives us information on the energy
balance of these regions. The main candidates for an explanation are
the absence of magnetic fields in the simulations and the fact that
the acoustic waves fed into the computational domain at the bottom
boundary do not include waves with frequencies above 20\,mHz.

The reason for the latter shortcoming is that the bottom boundary is
determined by an observed wave-field and high frequency waves are not
well determined observationally.  I first explore the possibility that
high frequency acoustic waves may account for the increased input and
address the issue of magnetic fields in the next section.

%fig: no-fit
%fig: Davos (Karlsen)
% Abbot & Hawley
% Allred et al

\subsection{High frequency waves}

% generation
% damping
% observations
% take text from lindau and apj_newenergy

Observationally it is difficult to detect high frequency acoustic
waves for two reasons: First, the seeing blurs the ground based
observations and makes these waves hard to observe.  Second, for both
ground based and space based observations the signal we get from high
frequency waves is weakened by the width of the response
function. \citet{mc-Wunnenberg2002} have summarized the various
attempts at detecting high frequency waves, and we refer to them for
further background.

Theoretically it is also non-trivial to determine the spectrum of
generated acoustic waves from convective motions. Analytic studies
indicate that there is a peak in the acoustic spectrum around periods
of 50\,s
\citep{mc-Musielak+Rosner+Stein+Ulmschneider1994,mc-Fawzy+Rammacher+Ulmschneider+Musielak+etal2002}
while results from high-resolution numerical simulations of convection
indicate decreasing power as a function of frequency
\citep{mc-Goldreich+Murray+Kumar1994,mc-Stein+Nordlund2001}.

Recently, \citet{mc-Fossum+Carlsson2005b} \&
\citet{mc-Fossum+Carlsson2006} analyzed observations from the
Transition Region And Coronal Explorer (TRACE) satellite in the
1600\,{\AA} passband. Simulations were used to get the width of the
response function \citep{mc-Fossum+Carlsson2005a} and to calibrate the
observed intensity fluctuations in terms of acoustic energy flux as
function of frequency at the response height (about 430\,km). It was
found that the acoustic energy flux at 430\,km is dominated by waves
close to the acoustic cut-off frequency and the high frequency waves
do not contribute enough to be a significant contributor to the
heating of the chromosphere.  Waves are detected up to 28\,mHz
frequency, and even assuming that all the signal at higher frequencies
is signal rather than noise, still gives an integrated energy flux of
less than 500 W\,m$^{-2}$, too small by a factor of ten to account for
the losses in the VAL3C model.  For the field free internetwork
regions used in the TRACE observations it is more appropriate to use
the VAL3A model that was constructed to fit the lowest intensities
observed with Skylab. It has about 2.2 times lower radiative losses
than VAL3C \citep{mc-Avrett1981} so there is still a major
discrepancy. One should also remember that
\citet{mc-Anderson+Athay1989} found three times higher energy
requirement than in the VAL3C model when they included the radiative
losses in millions of spectral lines, dominated by lines from
Fe\,II. As pointed out by \citet{mc-Fossum+Carlsson2006}, the main
uncertainty in the results is the limited spatial resolution of the
TRACE instrument (0.5\arcsec\ pixels corresponding to 1\arcsec\
resolution with a possible additional smearing from the little known
instrument PDF): ``There is possibly undetected wave power because of
the limited spatial resolution of the TRACE instrument.  The
wavelength of a 40 mHz acoustic wave is 180\,km and the horizontal
extent may be smaller than the TRACE resolution of
700\,km. Several arguments can be made as to why this effect is
probably not drastic. Firstly, 5 minute waves are typically
10--20\arcsec\ in coherence, 3 minute waves 5--10\arcsec. In both cases
3--6 times the vertical wavelength. This would correspond to close to
the resolution element for a 40\,mHz wave. Secondly, even a point
source excitation will give a spherical wave that will travel faster
in the deeper parts (because of the higher temperature) and therefore
the spherical wavefront will be refracted to a more planar wave. With
a distance of at least 500\,km from the excitation level it is hard to
imagine waves of much smaller extent than that at a height of 400
km. There is likely hidden power in the subresolution scales,
especially at high frequencies. Given the dominance of the low
frequencies in the integrated power, the effect on the total power
should be small. It is possible to quantify the missing power by
making artificial observations of 3D hydrodynamical simulations with
different resolution. This is not trivial since the results will be
dependent on how well the simuation describes the excitation of high
frequency waves and their subsequent propagation. Preliminary tests in
a 3D hydrodynamical simulation extending from the convection zone to
the corona \citep{mc-Hansteen2004} indicate that the effect of the
limited spatial resolution of TRACE on the total derived
acoustic power is below a factor of two. Although it is thus unlikely
that there is enough hidden subresolution acoustic power to provide
the heating for the chromosphere, the effect of limited spatial
resolution is the major uncertainty in the determination of the {\em
shape} of the acoustic spectrum at high frequencies.''

Another effect that goes in the opposite direction is that the
analysis assumes that {\em all} observed power above 5 mHz corresponds
to propagating acoustic waves. Especially at lower frequencies we will
also have a signal from the temporal evolotion of small scale
structures that in this analysis is mistakenly attributed to wave
power.

In a restrictive interpretation the result of
\citet{mc-Fossum+Carlsson2005b} is that acoustic heating can not
sustain a temperature structure like that in static, semi-empirical
models of the Sun.  Whether a dynamic model of the chromosphere can
explain the observations with acoustic heating alone has to be
answered by comparing observables from the hydrodynamic simulation
with observations. This was done by
\citet{mc-Wedemeyer-Bohm+Steiner+Bruls+Ramacher2007}. They come to the
conclusion that their dynamic model \citep{mc-Wedemeyer+etal2004} is
compatible with the TRACE observations (the limited spatial resolution
of the TRACE instrument severly affects the synthetic observations)
and that acoustic waves could provide enough heating of the
chromosphere. The synthetic TRACE images do not take into account
non-LTE effects or line opacities. It is also worth noting that the
model of \citet{mc-Wedemeyer+etal2004} does not have an average
temperature rise in the chromosphere and the dominant wave power is at
low frequencies close to the acoustic cut-off and not in the high
frequency part of the spectrum. Their study, however, is a major step
forward --- the question will have to be resolved by more realistic
modeling and synthesis of observations paired with high quality
observations.

The results by Fossum \& Carlsson also show that the neglect of
high-frequency waves in the simulations by Carlsson \& Stein is not an
important omission. Comparing their simulations with observations
shows that the agreement is good in the lower chromosphere
\citep{mc-Carlsson+Stein1997,mc-Judge+Carlsson+Stein2003} but lines
and continua formed above about 0.8\,Mm height have much too low mean
intensities. This is probably also true for the
\citet{mc-Wedemeyer+etal2004} model (since it has similar mean
temperature) but this needs to be checked by proper calculations.  It
thus seems inevitable that the energy balance in the middle and upper
chromosphere is dominated by processes related to the magnetic
field. This is consistent with the fact that the concept of a
non-magnetic chromosphere is at best valid in the low chromosphere ---
in the middle to upper chromosphere, the magnetic fields have spread
and fill the volume. Even in the photosphere, most of the area may be
filled with weak fields or with stronger fields with smaller filling
factor \citep{mc-Almeida2005,mc-Bueno+Shchukina+Ramos2004}.

\subsection{Comprehensive models in 3D}

% extension of 3D HD convection simulations to chromospheric heights
%  Skartlien

3D hydrodynamic simulations of solar convection have been very
successful in reproducing observations \citep[e.g.][]{mc-Nordlund1982,
mc-Stein+Nordlund1998,mc-Asplund+Nordlund+Trampedach+etal2000,mc-Voegler+Shelyag+Schussler+etal2005}.
It would be very natural to extend these simulations to chromospheric
layers to study the effect of acoustic waves on the structure,
dynamics and energetics of the chromosphere. This approach would then
include both the excitation of the waves by the turbulent motions in
the convection zone and their subsequent damping and dissipation in
chromospheric shocks.  For a realistic treatment there are several
complications. First, the approximation of LTE that works nicely in
the photosphere will overestimate the local coupling in chromospheric
layers. The strong lines that dominate the radiative coupling have a
source function that is dominated by scattering.  Second, shock
formation in the chromosphere makes it necessary to have a fine grid
or describe sub-grid physics with some shock capturing scheme.  Third,
it is important to take into account the long timescales for hydrogen
ionization/recombination for the proper evaluation of the energy
balance in the chromosphere
\citep{mc-Kneer1980,mc-Carlsson+Stein1992,mc-Carlsson+Stein2002}.

\citet{mc-Skartlien2000} addressed the first issue by extending the
multi-group opacity scheme of Nordlund to include the effects of
coherent scattering.  This modification made it possible to make the
first consistent 3D hydrodynamic simulations extending from the
convection zone to the chromosphere
\citep{mc-Skartlien+Stein+Nordlund2000}.  Due to the limited spatial
resolution, the emphasis was on the excitation of chromospheric wave
transients by collapsing granules and not on the detailed structure
and dynamics of the chromosphere.

%  Wedemeyer
%   3D HD

3D hydrodynamic simulations extending into the chromosphere with
higher spatial resolution were performed by
\citet{mc-Wedemeyer+etal2004}. They employed a much more schematic
description of the radiation (gray radiation) and did not include the
effect of scattering. This shortcoming will surely affect the amount
of radiative damping the waves undergo in the photosphere.  The
neglect of strong lines avoids the problem of too strong coupling with
the local conditions induced by the LTE approximation so in a way two
shortcomings partly balance out. The chromosphere in their simulations
is very dynamic and filamentary. Hot gas coexists with cool gas at all
heights and the gas is in the cool state a large fraction of the time.
As was the case in \citet{mc-Carlsson+Stein1995} they find that the
average gas temperature shows very little increase with height while
the radiation temperature does have a chromospheric rise similar to
the VAL3C model.  The temperature variations are very large, with
temperatures as low as 2000\,K and as high as 7000\,K at a height of 800
km. It is likely that the approximate treatment of the radiation
underestimates the amount of radiative damping thus leading to too
large an amplitude.

%   CO

The low temperatures in the simulations allow for a large amount of CO
to be present at chromospheric heights, consistent with observations.
For a proper calculation of CO concentrations it is important to take
into account the detailed chemistry of CO formation, including the
timescales of the reactions. This was done in 1D radiation
hydrodynamic models by
\citet{mc-AsensioRamos+TrujilloBueno+Carlsson+Cernicharo2003} and in
2D models by \citet{mc-Wedemeyer_CO2005}. The dynamic formation of CO
was also included in the 3D models and it was shown that CO-cooling
does {\em not} play an important role for the dynamic energy balance
at chromospheric heights \citep{mc-Wedemeyer+Steffen2007}.

% hydrogen ionization

Including long timescales for hydrogen ionization/recombination is
non-trivial.  In a 1D simulation it is still computationally feasible
to treat the full non-LTE problem in an implicit scheme (avoiding the
problem of stiff equations) as was shown by Carlsson \& Stein. The
same approach is not possible at present in 3D; the non-local coupling
is too expensive to calculate. Fortunately, the hydrogen ionization is
dominated by collisional excitation to the first excited level (local
process) followed by photo-ionization in the Balmer continuum.  Since
the radiation field in the Balmer continuum is set in the photosphere,
it is possible to describe the photoionization in the chromospheric
problem with a fixed radiation field (thus non-local but as a given
rate that does not change with the solution). This was shown to work
nicely in a 1D setting by \citet{mc-Sollum1999}.
\citet{mc-Leenarts+Wedemeyer2006} implemented the rate equations in 3D
but without the coupling back to the energy equation. The
non-equilibrium ionization of hydrogen has a dramatic effect on the
ionization balance of hydrogen in the chromosphere in their
simulation.

% including magnetic fields
%  Rosenthal et al
%  Carlsson & Stein 2002
%  Bogdan et al 2003
%  Carlsson & Bogdan 2006

Magnetic fields start to dominate over the plasma somehere in the
chromosphere.  Chromospheric plasma as seen in the center 
H$\alpha$ (e.g., \citet{mc-Rutten2007},
\citet{mc-DePontieu+Hansteen+Rouppe+etal2007}) is very clearly
organized along the magnetic structures. It is very likely that
acoustic heating alone is not sufficient to account for the radiative
losses in the chromosphere. It is thus of paramount importance to
include magnetic fields in chromospheric modeling but unfortunetely
the inclusion of magnetic fields increase the level of complexity
enormously. As was the case with acoustic waves, it is necessary to
perform numerical experiments and modeling in simplified cases in
order to fashion a basic physical foundation upon which to build our
understanding. A number of authors have studied various magnetic wave
modes and how they couple, see \citet{mc-Bogdan+etal2003} and
\citet{mc-Khomenko+Collados2006} for references.
\citet{mc-Rosenthal+etal2002} and \citet{mc-Bogdan+etal2003} reported
on 2D simulations in various magnetic field configurations in a
gravitationally stratified isothermal atmosphere, assuming an
adiabatic equation of state.  \citet{mc-Carlsson+Stein2002b} and
\citet{mc-Carlsson+Bogdan2006} reported on similar calculations in the
same isothermal atmosphere but this time in 3D and also studying the
effect of radiative damping of the shocks.
\citet{mc-Hasan+vanBallegooijen+Kalkofen+Steiner2005} studied the
dynamics of the solar magnetic network in two dimensions and
\citet{mc-Khomenko+Collados2006} studied the propagation of waves in
and close to a structure similar to a small sunspot.

The picture that emerges from these studies is that waves undergo mode
conversion, refraction and reflection at the height where the sound
speed equals the Alfv\'en speed (which is typically some place in the
chromosphere). The critical quantity for mode conversion is the angle
between the magnetic field and the k-vector: the attack angle. At
angles smaller than 30 degrees much of the acoustic, fast mode from
the photosphere is transmitted as an acoustic, slow mode propagating
along the field lines. At larger angles, most of the energy is
refracted/reflected and returns as a fast mode creating an
interference pattern between the upward and downward propagating
waves. When damping from shock dissipation and radiation is taken into
account, the waves in the low-mid chromosphere have mostly the
character of upward propagating acoustic waves and it is only close to
the reflecting layer we get similar amplitudes for the upward
propagating and refracted/reflected waves. It is clear that even
simple magnetic field geometries and simple incident waves create very
intricate interference patterns. In the chromosphere, where the wave
amplitude is expected to be large, it is crucial to include the
effects of the magnetic fields to understand the structure, dynamics
and energetics of the atmosphere. This is true even in areas
comparably free of magnetic field (such regions {\em may} exist in the
lower chromosphere).

% Seismology of the chromosphere

The fact that wave propagation is much affected by the magnetic field
topology in the chromosphere can be used for ``seismology'' of the
chromosphere.  Observational clues have been obtained by McIntosch and
co-workers: \citet{mc-McIntosh+etal2001} \&
\citet{mc-McIntosh+Judge2001} find a clear correlation between
observations of wave power in SOHO/SUMER observations and the magnetic
field topology as extrapolated from SOHO/MDI observations. These
results were extended to the finding of a direct correlation between
reduced oscillatory power in the 2D TRACE UV continuum observations
and the height of the magnetic canopy
\citep{mc-McIntosh+Fleck+Judge2003} and the authors suggest using
TRACE time-series data as a diagnostic of the plasma topography and
conditions in the mid-chromosphere through the signatures of the wave
modes present. Such helioseismic mapping of the magnetic canopy in the
solar chromosphere was performed by
\citet{mc-Finsterle+Jefferies+Cacciani+etal2004} and in a coronal hole
by \citet{mc-McIntosh+Fleck+Tarbell2004}.

The chromosphere is very inhomogeneous and dynamic. There is no simple
way of inverting the above observations to a consistent picture of the
chromospheric conditions. One will have to rely on comparisons with
full 3D Radiation-Magneto-Hydrodynamic forward
modeling. \citet{mc-Steiner+Vigeesh+Krieger+etal2007} tracked a
plane-parallel, monochromatic wave propagating through a
non-stationary, realistic atmosphere, from the convection-zone through
the photosphere into the magnetically dominated chromosphere. They
find that a travel time analysis, like the ones mentioned above,
indeed is correlated with the magnetic topography and that high
frequency waves can be used to extract information on the magnetic
canopy.

% Piecing it all together
%  Hansteen 2004
%  

\begin{figure}
  \centering
  \includegraphics[width=\textwidth]{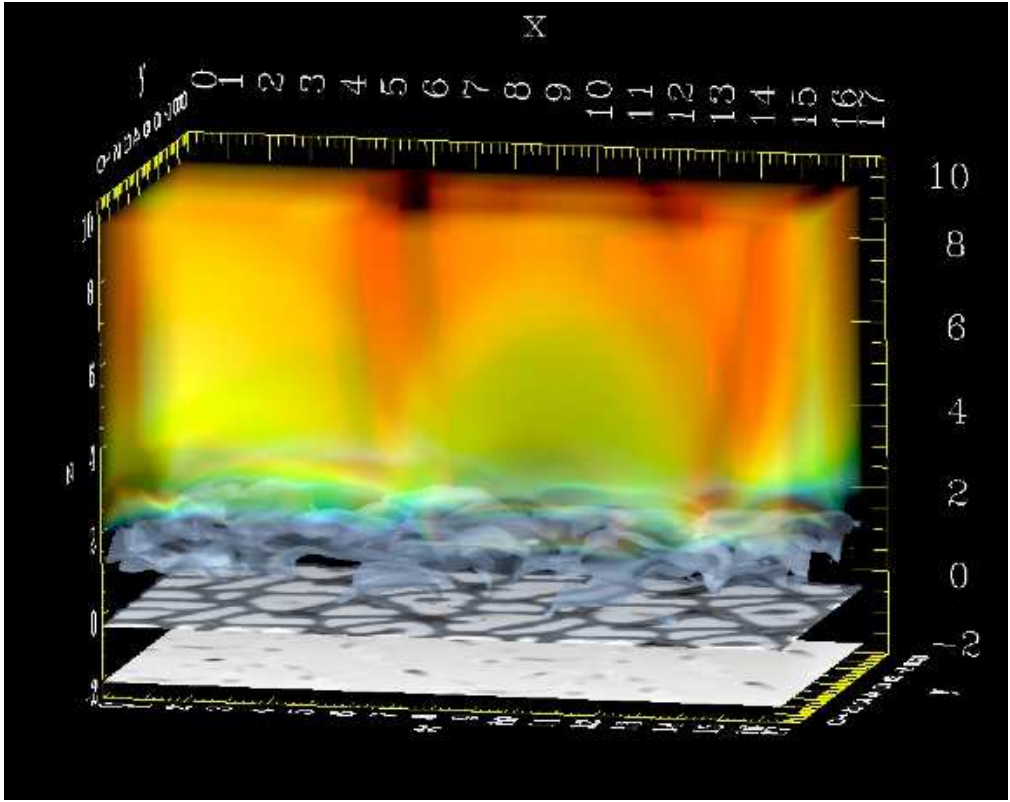}
  \caption[]{\label{mc-fig:simulation}
    Temperature structure in a 3D simulation box. The bottom plane shows
    the temperature at 1.5 Mm below $\tau_{500}$=1 ranging from 
    $15\,700$\,K in down-flowing plumes to $16\,500$ in
    the gas flowing into the simulation domain. The next plane is in
    the photosphere and shows hot granules and cool intergranular lanes.
    In the chromosphere the isothermal surfaces show pronounced small
    scall structures and corrugated shock fronts. 
    The upper 8\,Mm is filled with plasma at transition region and coronal
    temperatures up to 1\,MK.
}\end{figure}

Including not only 3D hydrodynamics but in addition the magnetic
field, and extending the computational domain to include the corona is
a daunting task. However, the development of modern codes and
computational power is such that it is a task that is within reach of
fulfillment.  \citet{mc-Hansteen2004} reported on the first results
from such comprehensive modeling. The 3D computational box is
$16\times8\times12$\,Mm in size extending 2\,Mm below and 10\,Mm above the
photosphere. Radiation is treated in detail, using multi-group
opacities including the effect of scattering \citep{mc-Skartlien2000},
conduction along field-lines is solved for implicitly and optically
thin losses are included in the transition region and corona. For a
snapshot of such a simulation, see Fig.\ref{mc-fig:simulation}.

After a relaxation phase from
the initial conditions, coronal temperatures are maintained
self-consistently by the injection of Pointing flux from the convective buffeting of
the magnetic field, much as in the seminal simulations by Gudiksen \&
Nordlund (2002, 2005a, 2005b).
\nocite{mc-Gudiksen+Nordlund2002}
\nocite{mc-Gudiksen+Nordlund2005a}
\nocite{mc-Gudiksen+Nordlund2005b}

It is clear from these simulations that the presence of the magnetic
field has fundamental importance for chromospheric dynamics and the
propagation of waves through the chromosphere. It is also clear that
magnetic fields play a role in the heating of the chromosphere
\citep{mc-Hansteen+Carlsson+Gudiksen2007}.

% Future prospects

There are several hotly debated topics in chromospheric modeling
today: Is the internetwork chromosphere wholly dynamic in nature or
are the dynamic variations only minor perturbations on a semi-static
state similar to the state in semi-empirical models
\citep[e.g.,][]{mc-Kalkofen+Ulmschneider+Avrett1999}?  Is there a
semi-permanent cold chromosphere (where CO lines originate) or is the
CO just formed in the cool phases of a dynamic atmosphere? Is there
enough chromospheric heating in high frequency waves of small enough
spatial extent that they are not detected by the limited spatial
resolution of TRACE?  What is the role of the magnetic field (mode
conversion of waves, reconection, currents, channeling of waves)?  A
reason for conflicting results is the incompleteness of the physical
description in the modeling and the lack of details (spatial and
temporal resolution) in the observations. We are rapidly progressing
towards the resolution of this situation. New exciting observations at
high temporal and spatial resolution (especially from the Swedish 1-m
Solar Telescope on La Palma) are changing our view of the
chromosphere.  New observing facilities are on the verge of coming on
line (GREGOR, Hinode). On the modeling side, several groups have
developed codes that start to include the most important ingredients
for a comprehensive modeling of the dynamic chromosphere
\citep[e.g.,][]{mc-Hansteen2004,
mc-Schaffenberger+Wedemeyer+Steiner+Freytag2006}.  There is still more
work to do with simulations of idealized cases to build up a
foundation for our understanding and this is also a field with several
groups active at present. The future for chromospheric modeling thus
looks both promising and exciting.\\

%!! reverted to version before fill paragraph had put "command." into text
\acknowledgements %%% Text of acknowledgements runs on after this command.  
This work was supported by the Research Council of Norway
grant 146467/420 and a grant of computing time from the Program for
Supercomputing.

%% References via BibTeX
%%%%%%%%%%%%%%%%%%%%%%%%%%%%%%%%%%%%%%%%%%%%%%%%%%%%%%%%%%%%%%%%%%%%%%%%%%%%
%RR Use the following two commands to generate the bibliography
%RR automatically with BibTeX, and then insert the resulting .bbl file,
%RR or collect \bibitem info for each paper from ADS (``preferred format'')

%\bibliographystyle{cspm-bib}          %RR copy of my old aabib.bst
%\bibliography{aajour,finstruc,rotwaves,modcont,mc}

\end{document}